\documentclass[aps,prl,twocolumn,groupedaddress]{revtex4-1}

\usepackage{latexsym}
\usepackage[T1]{fontenc}
\usepackage{times}
\usepackage{graphicx}
\usepackage{amsmath}

\usepackage[colorlinks=true,bookmarks=false,citecolor=blue,urlcolor=blue,pdfstartview=FitH]{hyperref}

\newcommand{\dr}{\mathrm{d}r}
\newcommand{\Tes}{T^{\textrm{\scriptsize es}}}

\begin{document}

\title{Electrostriction and guidance of sound by light in optical fibers}

\author{Jean-Charles Beugnot and Vincent Laude}
\email{v.laude@femto-st.fr}
\affiliation{Institut FEMTO-ST, Universit\'e de Franche-Comt{\'e} and CNRS, Besan\c{c}on, France}

\date{\today}

\begin{abstract}
We investigate the generation of phonon wavepackets in optical fibers via electrostriction from coherent optical waves.
Solving the elastodynamic equation subject to the electrostrictive force, we are able to reproduce experimental spectra found in standard and photonic crystal fibers.
We discuss the two important practical cases of forward interaction, dominated by elastic resonances of the fiber, and backward interaction, for which an efficient mechanism of phonon guidance by light is found.
The latter result describes the formation of the coherent phonon wavepacket involved in stimulated Brillouin scattering.
\end{abstract}


\maketitle

Electrostriction (ES) describes the coherent generation of acoustic phonons from the interference of two frequency-detuned optical waves.
The effect is closely related to Brillouin light scattering (BLS), or scattering of an incident photon by an acoustic phonon of a solid material, accompanied by a frequency shift of the scattered photon. 
The combination of BLS and ES explains stimulated Brillouin scattering (SBS), an efficient three-wave interaction involving two optical waves and a phonon wavepacket satisfying both energy and momentum conservation. SBS is a fundamental limiting phenomenon for signal transport over optical fibers~\cite{boydBOOK2008}. Interactions of photons and acoustic resonances of photonic crystal fibers (PCF)~\cite{kangPRL2010} and optical waveguides~\cite{rakichPRX2012} have received much attention, for instance in view of obtaining strong optoacoustic interactions~\cite{kangNP2009} or of reducing noise in quantum optics experiments~\cite{elserPRL2006}.
ES in optical fibers is not easily observed, since while the incident and the scattered optical waves are directly accessible to experiment, the intervening acoustic phonons are only witnessed via their induced effects on light. As a result, most theories eliminate phonon dynamics to concentrate on the time evolution of the amplitudes of optical waves. This process, however, can only be performed under strong assumptions on the nature of the phonon wavepackets generated by ES. Guided acoustic wave Brillouin scattering (GAWBS) theories, for instance, assume the acoustic modal shapes and resonant frequencies to be known beforehand, so that scattering cross-sections can be computed from overlap integrals~\cite{shelbyPRB1985,daineseNP2006,beugnotOL2007}.
Modal approaches, however, give no insight into the mechanism of selection of particular phonons from the electrostrictive force.

In this Letter, we propose a direct calculation of phonon wavepackets induced by electrostriction in optical fibers, basing only on knowledge of the fiber geometry and of the particular incident guided optical waves. Specifically, an elastodynamic equation is solved for a fiber with an arbitrary cross-section subject to an optical force. By comparing with experimental spectra obtained with standard optical fiber and photonic crystal fiber with a solid core, we show that the computed phonon wavepackets as a function of detuning frequency explain both backward and forward ES and provide an estimation of the efficiency of the process. Under the phase-matching conditions of SBS, furthermore, our results imply that the phonon wavepacket generated by electrostriction is actually guided by light.

Standard SBS models for optical fibers are based on a rather strong plane-wave approximation~\cite{shenPR1965,krollJAP1965} and on the picture of ES-generated pressure or density fluctuation waves~\cite{boydBOOK2008}.
In contrast to this simple picture, theory and experiment have shown the relevance of including both shear and longitudinal elastic waves together with a precise description of the structure~\cite{shelbyPRB1985,laudePRB2005,daineseNP2006,kangPRL2010}.
Let us consider an optical fiber with an arbitrary cross-section.
We assume the total incident optical field results from the superposition of two frequency-detuned guided modes
\begin{equation}
E(r,z;t) = E^{(1)}(r) e^{i(\omega_1 t - k_1 z)} + E^{(2)}(r) e^{i(\omega_2 t - k_2 z)} ,
\end{equation} 
with angular frequencies $\omega_{1(2)}$ and axial wavevectors $k_{1(2)}$.
This optical distribution induces an optical force given by divergence of the symmetric ES stress tensor $T_{ij}^{es} =  -\epsilon_0 \chi_{klij} E_{k} E^{*}_{l}$, with the rank-4 susceptibility tensor $\chi_{klij}=\varepsilon_{km} \varepsilon_{ln}  p_{mnij}$ and $p_{mnij}$ the elastooptic tensor. $\epsilon_0$ is the permitivity of vacuum.
The force term with detuning frequency $\omega = \omega_1 - \omega_2$ is proportional to $E^{(1)}_{k} E^{(2)*}_{l}\exp(i(\omega t - k z))$ with $k = k_1 - k_2$.
Because we are considering the generation of elastic waves with low frequency with respect to optical frequencies, $\omega \ll \omega_{1,2}$.
If the two optical waves are propagating in the same direction, a situation typical of GAWBS~\cite{shelbyPRB1985}, $k \approx 0$ and we speak of forward ES.
If they are propagating in opposite directions, which is the SBS situation,  $k \approx 2 k_1$ and we speak of backward ES.
Because of the particular source term, the following \textit{ansatz} is assumed for the displacements of the ES-generated elastic wave 
\begin{equation}
u_{i}(r,z;t) = \bar{u}_{i}(r) e^{i(\omega t - k z)} ,
\end{equation}
where the transverse dependence $\bar{u}_{i}(r)$ is the unknown of the model.
The elastodynamic equation
\begin{equation}
\rho \frac{\partial^2 u_i}{\partial t^2} - (c_{ijkl} u_{k,l})_{,j}  = - T_{ij,j}^{es} ,
\label{eq3}
\end{equation}
with $c_{ijkl}$ the rank-4 tensor of elastic constants, is written in variational form by left-multiplying by virtual displacement field $v_i$ and integrating over the cross-section, $S$.
Further applying Green's theorem, we get
\begin{equation}
- \omega^2 \int_S \rho v^*_i u_i + \int_S v^*_{i,j} c_{ijkl} u_{k,l} = \int_S \dr v^*_{i,j} T^{\textrm{\scriptsize es}}_{ij} ,
\end{equation}
which amounts to the theorem of virtual work for the ES force.
For practical computations, the Galerkin nodal finite element method (FEM) is then employed to transform the integral equation into the linear system
\begin{equation}
(K(k) - \omega^2 M) U = X(k) T^{\textrm{\scriptsize es}} ,
\label{eq5}
\end{equation}
with mass matrix $M$, stiffness matrix $K(k) = K_0 + k K_1 + k^2 K_2$, and $X(k)=X_0 + k X_1$.
$U$ is the vector of nodal displacements $\bar{u}_i$ and $T^{\textrm{\scriptsize es}}$ is the vector of electrostriction stress tensor value at nodal points, $\bar{T}_{ij}^{es} =  -\epsilon_0 \chi_{klij} E^{(1)}_{k} E^{(2)*}_{l}$.
Equation~(\ref{eq5}) is our central result. Its solution as a function of frequency detuning gives the rigorous distribution of displacements within the waveguide cross-section and can be directly compared to experiment, as we show next. The ES stress tensor is uniquely defined by the optical modal distribution. Matrices $K(k)$ and $X(k)$ are polynomials in the wavevector detuning $k$, thus unifying in the same formula forward and backward ES, as well as any intermediate situation.

As a test for our model, we selected a 1 km long single mode fiber (SMF28) and a 400 m long large core PCF with similar optical propagation properties.
Single mode guidance is achieved by total internal reflection: SMF28 has a Ge-doped core with larger refractive index than surrounding undoped silica, while PCF has a pure-silica core surrounded by a holey structure with lower effective index.
The silica PCF shown as an inset in Fig.~\ref{fig2}(a) has a hole diameter of 4.6 $\mu$m and an air filling ratio $d / \Lambda =$ 0.58, resulting for the fundamental mode in effective index $n_{\rm{eff}}=1.441$ and optical effective area $A_{\rm{eff}} = 70$ $\mu$m$^2$.  
The SMF28 fiber has $n_{\rm{eff}}=1.446$ and $A_{\rm{eff}} = 78.3$ $\mu$m$^2$, while the core diameter is $8.2~\mu$m.

\begin{figure}
\centering
\includegraphics[width=80mm]{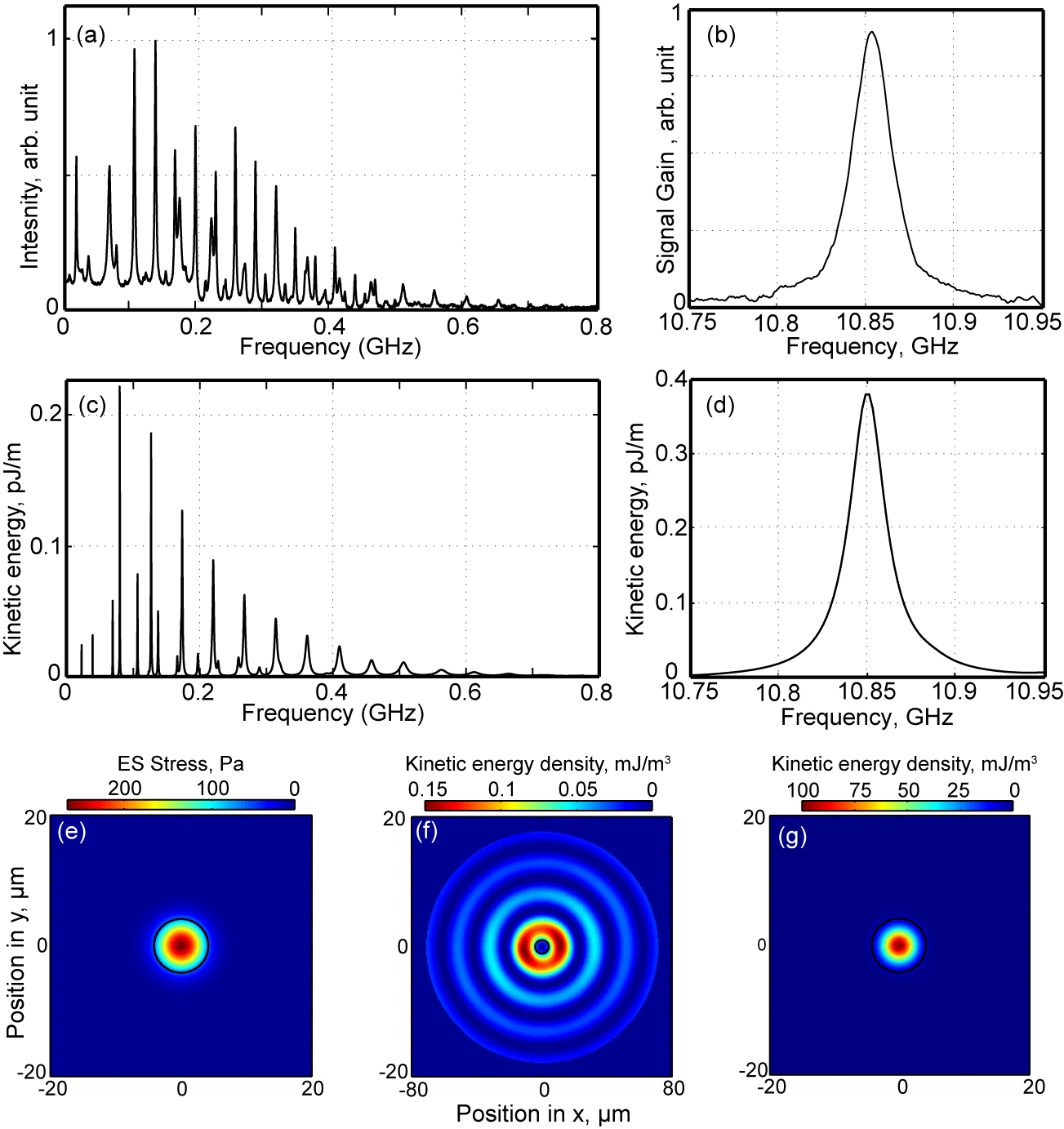}
\caption{Comparison between experiment and ES model in single mode fiber (SMF28).
(a) Experimental and (c) numerical forward spectra ($k=0$).
(b) Experimental and (d) numerical backward spectra ($k=2k_1$).
(e) Spatial distribution of the ES force generated by the fundamental optical mode at a wavelength of 1550~nm with 1 W optical power. The black circle limits the germanium doping core area.
(f,g) Computed kinetic energy density of phonon wavepackets at 178 MHz (forward ES) and 10.85 GHz (backward ES).}
\label{fig1}
\end{figure}

ES-generated phonons inside an optical fiber are hardly observable directly.
They can, however, be observed indirectly in a manner similar to spontaneous Brillouin gain spectrum measurement or Brillouin sensing, by monitoring light diffracted from a coherent pump wave.
A heterodyne detection experimental setup was used for investigating backward ES~\cite{beugnotOE2007}.
To observe forward ES, we used a fiber loop mirror interferometer~\cite{beugnotOL2007}.
The measured spectra for the SMF28 fiber are show in Fig.~\ref{fig1}(a,b). The forward spectrum has many sharp peaks that gradually vanish for frequencies above 600 MHz.
The peaks are known to be directly related to elastic resonances of the fiber.
The backward spectrum shows a single Lorentzian peak with a linewidth of 27 MHz and a central frequency $10.85$ GHz.
The PCF case shown in Fig.~\ref{fig2}(a,b) looks similar, though resonance frequencies are slightly different.
Forward ES resonances in this case are known to be related to the holey structure rather than to the external cladding of the fiber.

\begin{figure}
\centering
\includegraphics[width=80mm]{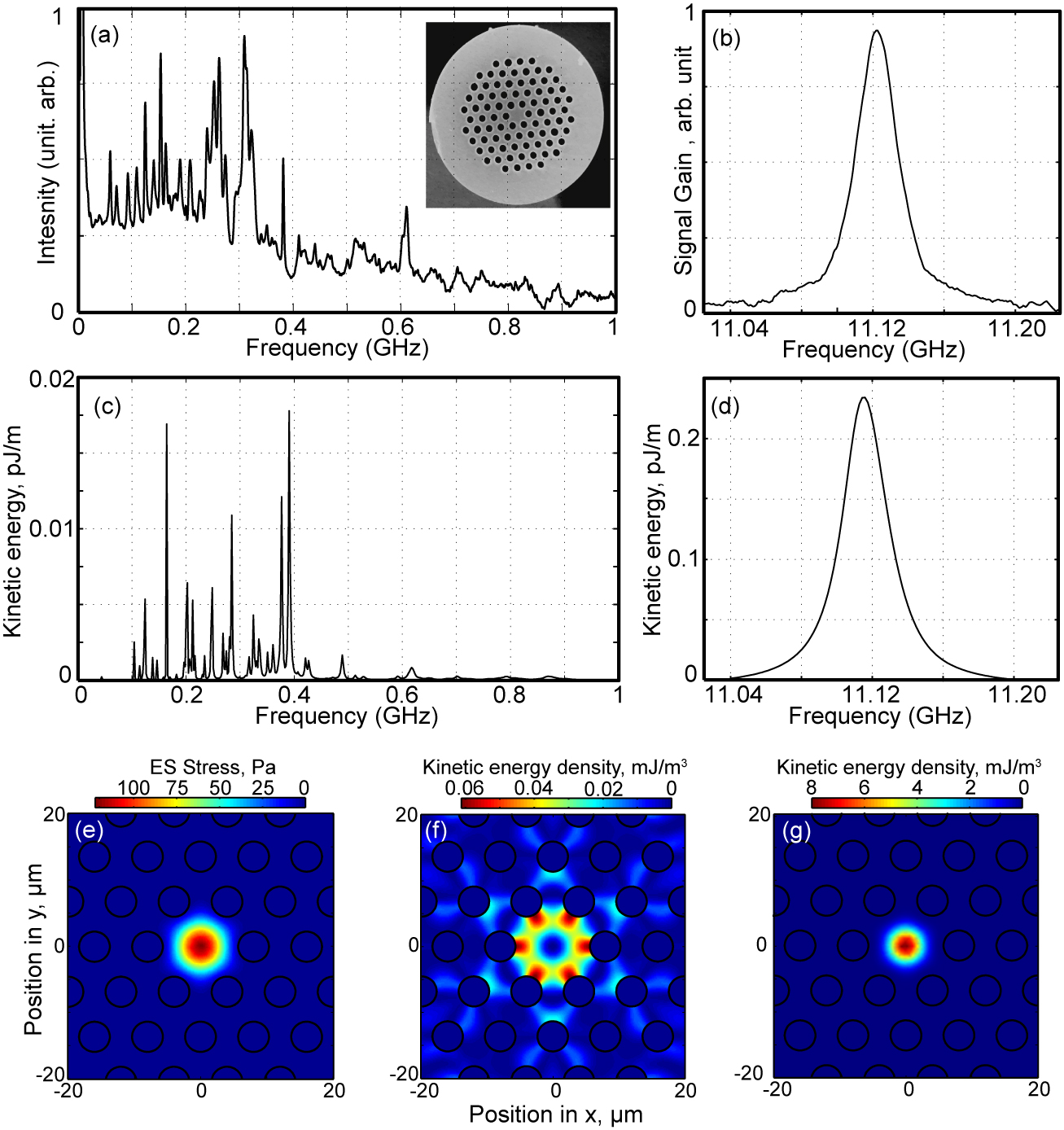}
\caption{Comparison between experiment and ES model in large core PCF.
(a) Experimental and (c) numerical forward spectra ($k=0$). The inset in (a) shows a cross-section of the fiber.
(b) Experimental and (d) numerical backward spectra ($k=2k_1$).
(e) Spatial distribution of the ES force generated by the fundamental optical mode at a wavelength of 1550 nm with 1 W optical power.
(f,g) Computed kinetic energy density of phonon wavepackets at 390 MHz (forward ES) and 11.01 GHz (backward ES).}
\label{fig2}
\end{figure}

\begin{table}[!t]
\caption{Independent material constants for isotropic silica~\cite{royerBOOK1999}. Values for 0.36\%-GeO$_2$-doped silica given in parenthesis are deduced from \cite{koyamadaJOLT2004}.}
\label{tab1}
\begin{tabular}{llll}
\hline
\small
Elast. const., GPa & $c_{11}=78$ ($76$) & $c_{12}=16$ ($16.15$) & $c_{44}=31$ ($29.9$) \\
Photoelast. const. & $p_{11}=0.12$ & $p_{12}=0.27$ & $p_{44}=-0.073$ \\
Mat. dens., kg/m$^3$ & \multicolumn{3}{l}{$\rho=2203$ ($2254$)} \\
Refractive index & \multicolumn{3}{l}{$n=1.444$ ($1.4492$) at $\lambda=1550$ nm} \\
\hline
\end{tabular}
\end{table}

We then proceeded to model experimental results using Eq.~(\ref{eq5}).
A FEM mesh covering the full fiber cross-section was first created, with the outer boundaries left free.
The mesh of the PCF was extracted from the scanning electron microscope image shown as an inset in Fig.~\ref{fig2}(b).
Fundamental optical modes are first obtained and normalized for unit transported power.
ES stress distributions are displayed in Figs.~\ref{fig1}(e) and \ref{fig2}(e).
For both fibers, the ES stress is confined in the core roughly as the square of the optical field.
Next, the ES-driven elastodynamic equation was solved for the displacement of the elastic wave by imposing the phonon wave vector $k$ and scanning the detuning frequency $\omega$.
Independent material constants considered for silica are shown in Table~\ref{tab1}.
To include the phonon lifetime, elastic losses were incorporated in the ES model by considering a complex elastic tensor whose imaginary part is a constant viscosity tensor times frequency~\cite{moiseyenkoPRB2011}. This loss model is compatible with the usual assumption that the $Qf$ product is a constant for a given material, with $Q$ the quality factor and $f$ the frequency. This is the only adjustable parameter in the model. The value $Qf = 5.10^{12}$ Hz was selected to fit the SBS linewidth of SMF28 at 11.01 GHz.

Calculated forward ($k=0$) and backward ($k=2k_1$) ES spectra are shown in Fig.~\ref{fig1} and Fig.~\ref{fig2} for SMF28 and PCF, respectively.
These spectra are obtained by evaluating the kinetic energy in the phonon wavepacket, $\omega^2 / 2 \int_S \rho |u_i|^2$, as a function of frequency detuning.
The agreement with experimental spectra is excellent in the backward case, with only one Lorentzian peak appearing.
In the forward case, the appearance of a sequence of sharp resonance peaks and its gradual disapearance with increasing frequency are correctly reproduced, though relative peak heights are not accurately predicted.
We display in Fig.~\ref{fig1}(f) and Fig.~\ref{fig2}(f) the kinetic energy densities at resonance for selected frequencies in the forward ES spectrum.
For SMF28, the energy distribution fills the whole fiber.
For PCF, in contrast, the phonon wavepacket is confined by the holey microstructure.
For both fibers, the maximum displacement is of the order of 150 fm and is dominantly transversal.
The situation is dramatically different in the backward case, since the phonon wavepacket is found to be confined to the core for both fibers.
For both fibers, the maximum displacement is of the order of 50 fm and is dominantly longitudinal, though transversal displacements don't vanish.

Let us discuss the above results in the light of elastic normal modes of the fiber section.
Such an approach has been successful to describe GAWBS in classical fibers~\cite{ThomasPRB1979} and recently in PCF~\cite{daineseOE2006,beugnotOL2007,wiederheckerPRL2008,kangPRL2010}.
It has also been used to describe the phononic crystal properties of PCF~\cite{laudePRB2005,daineseNP2006}.
For a given wavevector $k$, modes are obtained in the absence of any applied force as solutions of the eigenvalue problem $K U = \omega^2 M U$.
Because the fiber cross-section has finite dimensions, the spectrum of eigenmodes $\omega_n$ is discrete.
Modes are such that their potential and kinetic energy are equal, $U_n^\dag K U_n = \omega_n^2 U_n^\dag M U_n$.
They are furthermore orthogonal, or $U_m^\dag K U_n = U_m^\dag M U_n = 0$, provided $\omega_m \neq \omega_n$.
They thus constitute a complete basis, and the solution to the forced equation can be expanded as
\begin{equation}
U = \sum_n \frac{\omega_n^2}{\omega_n^2 - \omega^2} \frac{U_n^\dag X T^{\scriptsize (es)}}{U_n^\dag K U_n} U_n .
\end{equation}
As a consequence, the phonon response to the electrostrictive excitation is a sequence of Lorentzian-shaped resonances; at each resonance frequency, the modal shape of the phonon wavepacket is a normal mode of the fiber.
Clearly, the electrostriction gain is governed by the overlap integral $U_n^\dag X T^{\scriptsize (es)} = \int_V \dr (u_n)^*_{i,j} T^{\textrm{\scriptsize es}}_{ij}$, that compares the strain field of each mode with the electrostrictive stress distribution.
Because of elastic losses, the response cannot diverge to infinity at resonance.
Writing $\omega_n = \omega_{rn} + \imath \omega_{in}$,
\begin{equation}
U(\omega=\omega_{rn}) \approx \frac{\omega_{rn}}{2 \imath \omega_{in}} \frac{U_n^\dag X T^{\scriptsize (es)}}{U_n^\dag K U_n} U_n .
\end{equation}
The first factor in this expression is proportional to the Q-factor of the resonance.
As the Q-factor in our viscoelastic loss model scales with the inverse of frequency, it is clear that resonance peaks must decrease with increasing frequency, as observed experimentally and with the direct computation in the forward ES case.

\begin{figure}
\centering
\includegraphics[width=80mm]{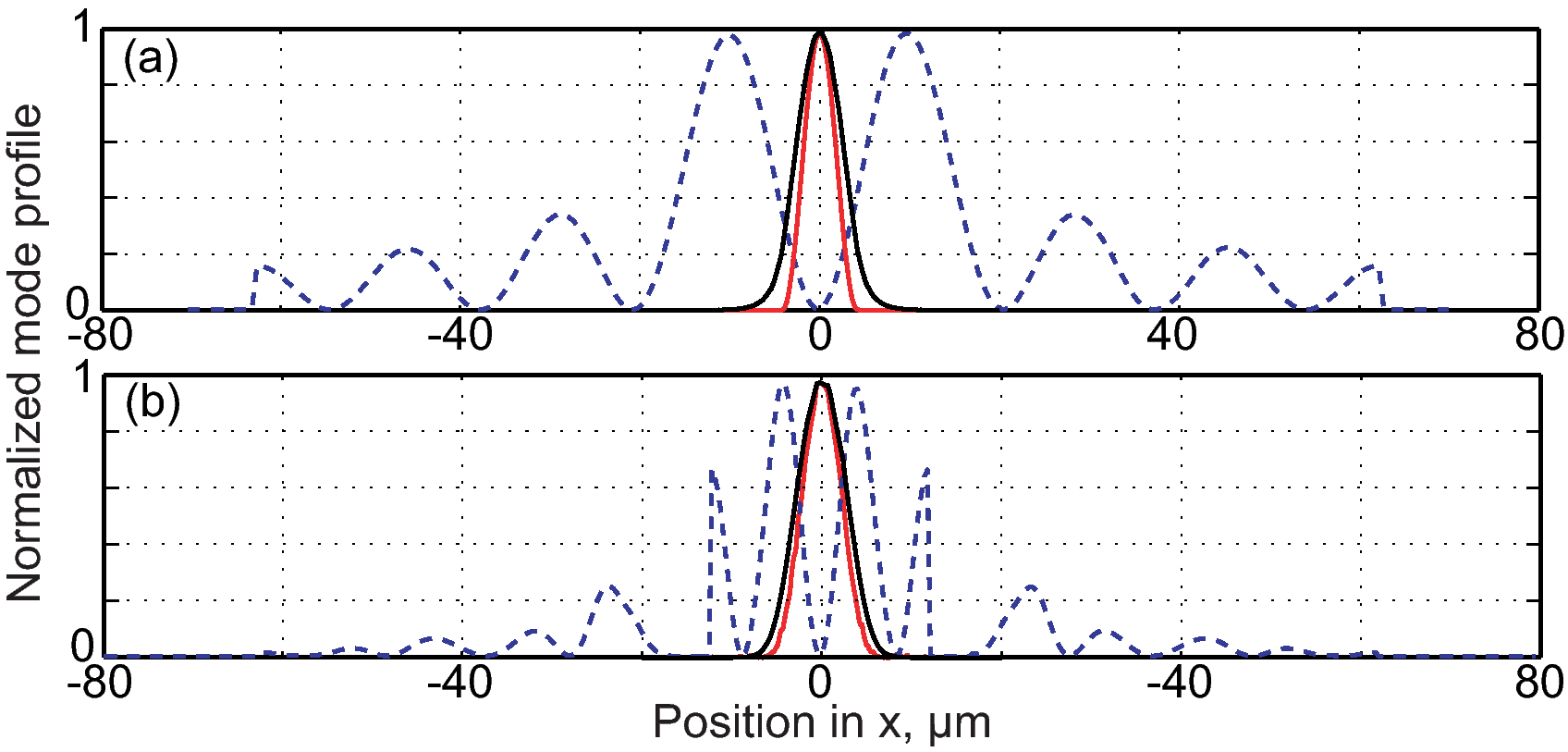}
\caption{Normalized X-profiles of optical power density and kinetic energy density in (a) photonic crystal fiber and (b) standard fiber. The black line shows the fundamental optical mode guided by total internal reflection. The dashed blue line shows the most efficient elastic mode in the forward ES case and the red line shows the phonon wavepacket excited in the backward ES case.}
\label{fig5}
\end{figure}

Now it is clear that the above normal mode argument cannot explain the observations for backward ES ($k=2k_1$).
Actually, as $k$ becomes large, there exist more and more normal modes within any frequency interval, and the above analysis in terms of a discrete spectrum of modes becomes inadequate.
The backward case can instead be treated under the following short-wavelength approximation.
For an optical wavelength of 1.55 $\mu$m in a vacuum, indeed, the optical wavelength in silica is about 1 $\mu$m and the acoustic phonon wavelength is thus about 500 nm. This is typically much smaller than the diameter of a classical optical fiber, but still comparable to the size of the core of some PCF or integrated optical waveguides.
We consider specifically core diameters exceeding a few microns in the following discussion.
Neglecting $K_0$ and $k K_1$ in the stiffness matrix expression, we have
\begin{equation}
k^2 K_2 U_B = \omega^2_B M U_B ,
\label{eq8}
\end{equation}
which defines the usual Brillouin frequency $\omega_B$.
$U_B$, the eigenvector corresponding to eigenfrequency $\omega_B$, is a simple plane wave~\footnote{The eigenvector is actually exactly a plane wave if the material constituting the waveguide is homogeneous, in which case Eq. (\ref{eq8}) is simply the Cristofell equation for elastic waves in solids}.
In the short-wavelength approximation, $X \approx k X_1$ and the electrostrictive force is mostly longitudinal because of the particular symmetries of the elasto-optic tensor of silica.
As a consequence, shear waves satisfying Eq. (\ref{eq8}) will hardly be excited and we can take $U_B$ as dominantly longitudinally polarized, and the velocity $\omega_B/k_B$ as very close to the velocity for longitudinal waves in silica, $\sqrt{c_{11}/\rho}=5950$ m/s.
It must be stressed that this situation is peculiar to silica, and would not necessarily apply to other materials, such as silicon.
We further observe that because silica is isotropic, the solution to Eq. (\ref{eq8}) is actually independent of the propagation direction.
The actual response of the medium to the ES stress, $U$, can then be decomposed as an angular spectrum of plane waves, satisfying $k^2 K_2 U = \omega^2_B M U$ to a good approximation for $\omega \approx \omega_B$.
Replacing this relation in Eq. (\ref{eq5}), we obtain that
\begin{equation}
U \approx [K_0 + k_B K_1 + (\omega^2_B - \omega^2)M]^{-1} X \Tes .
\end{equation}
This expression represents a Lorentzian distribution centered on the Brillouin frequency.
Close to resonance, the term $(\omega^2_B - \omega^2)$ is minimum, but non vanishing because of phonon loss.
Significantly, the transverse distribution of the phonon wavepacket is similar to the electrostrictive stress distribution, $\Tes$.
This is very different from the forward case, for which this distribution is given by the normal modes of the fiber, and thus extends throughout the fiber cross-section.
Fig.~\ref{fig5} illustrates this point by showing profiles of the distribution of the phonon wavepackets.
We believe this particular distribution of the phonon wavepacket explains why SBS is remarkably efficient in optical fibers, independently of the fact that the involved acoustic phonons are guided or not by the transversal structure of the fiber.

As a conclusion, the consideration of electrostrictive forces has allowed us to evaluate precisely the phonon wavepackets generated in optical fibers by Brillouin-type interactions.
Our model encompasses the forward interaction case, including GAWBS, and the backward interaction case, including SBS.
In all expressions we have derived, the frequency response to electrostriction is a Lorentzian function whose width is dictated by intrinsic phonon losses, or the inverse of the phonon life time, possibly dependent on frequency.
Significantly, the phonon wavepacket generated via backward ES is naturally guided by the light that gave it birth.
Our method could furthermore be extended to almost any type of optical waveguide.
Silicon nanostructures, for instance, have attracted attention recently in view of obtaining optomechanical coupling in structures supporting simultaneous confinement of elastic and optical waves~\cite{vantourhoutNP2010}. Optical forces, including ES and radiation pressure~\cite{kippenbergS2008}, have been predicted to scale to large values in nanoscale waveguides \cite{rakichPRX2012}. The model we have presented, possibly supplemented with optical surface force terms, should yield the phonon wavepackets generated in such cases.

\begin{acknowledgments}
We thank T. Sylvestre and H. Maillotte for fruitful discussions and acknowledge financial support from the European program INTERREG IV (CD-FOM) and the European Community\textquoteright s Seventh Framework program (FP7/2007-2013) under grant agreement number 233883 (TAILPHOX).
\end{acknowledgments}

\bibliography{vince}

\begin{thebibliography}{21}%
\makeatletter
\providecommand \@ifxundefined [1]{%
 \@ifx{#1\undefined}
}%
\providecommand \@ifnum [1]{%
 \ifnum #1\expandafter \@firstoftwo
 \else \expandafter \@secondoftwo
 \fi
}%
\providecommand \@ifx [1]{%
 \ifx #1\expandafter \@firstoftwo
 \else \expandafter \@secondoftwo
 \fi
}%
\providecommand \natexlab [1]{#1}%
\providecommand \enquote  [1]{``#1''}%
\providecommand \bibnamefont  [1]{#1}%
\providecommand \bibfnamefont [1]{#1}%
\providecommand \citenamefont [1]{#1}%
\providecommand \href@noop [0]{\@secondoftwo}%
\providecommand \href [0]{\begingroup \@sanitize@url \@href}%
\providecommand \@href[1]{\@@startlink{#1}\@@href}%
\providecommand \@@href[1]{\endgroup#1\@@endlink}%
\providecommand \@sanitize@url [0]{\catcode `\\12\catcode `\$12\catcode
  `\&12\catcode `\#12\catcode `\^12\catcode `\_12\catcode `\%12\relax}%
\providecommand \@@startlink[1]{}%
\providecommand \@@endlink[0]{}%
\providecommand \url  [0]{\begingroup\@sanitize@url \@url }%
\providecommand \@url [1]{\endgroup\@href {#1}{\urlprefix }}%
\providecommand \urlprefix  [0]{URL }%
\providecommand \Eprint [0]{\href }%
\providecommand \doibase [0]{http://dx.doi.org/}%
\providecommand \selectlanguage [0]{\@gobble}%
\providecommand \bibinfo  [0]{\@secondoftwo}%
\providecommand \bibfield  [0]{\@secondoftwo}%
\providecommand \translation [1]{[#1]}%
\providecommand \BibitemOpen [0]{}%
\providecommand \bibitemStop [0]{}%
\providecommand \bibitemNoStop [0]{.\EOS\space}%
\providecommand \EOS [0]{\spacefactor3000\relax}%
\providecommand \BibitemShut  [1]{\csname bibitem#1\endcsname}%
\let\auto@bib@innerbib\@empty
\bibitem [{\citenamefont {Boyd}(2008)}]{boydBOOK2008}%
  \BibitemOpen
  \bibfield  {author} {\bibinfo {author} {\bibfnamefont {R.~W.}\ \bibnamefont
  {Boyd}},\ }\href@noop {} {\emph {\bibinfo {title} {Nonlinear Optics}}},\
  \bibinfo {edition} {3rd}\ ed.\ (\bibinfo  {publisher} {Academic Press},\
  \bibinfo {year} {2008})\BibitemShut {NoStop}%
\bibitem [{\citenamefont {Kang}\ \emph {et~al.}(2010)\citenamefont {Kang},
  \citenamefont {Brenn},\ and\ \citenamefont {Russell}}]{kangPRL2010}%
  \BibitemOpen
  \bibfield  {author} {\bibinfo {author} {\bibfnamefont {M.~S.}\ \bibnamefont
  {Kang}}, \bibinfo {author} {\bibfnamefont {A.}~\bibnamefont {Brenn}}, \ and\
  \bibinfo {author} {\bibfnamefont {P.~St.~J.}\ \bibnamefont {Russell}},\ }\href
  {\doibase 10.1103/PhysRevLett.105.153901} {\bibfield  {journal} {\bibinfo
  {journal} {Phys. Rev. Lett.}\ }\textbf {\bibinfo {volume} {105}},\ \bibinfo
  {pages} {153901} (\bibinfo {year} {2010})}\BibitemShut {NoStop}%
\bibitem [{\citenamefont {Rakich}\ \emph {et~al.}(2012)\citenamefont {Rakich},
  \citenamefont {Reinke}, \citenamefont {Camacho}, \citenamefont {Davids},\
  and\ \citenamefont {Wang}}]{rakichPRX2012}%
  \BibitemOpen
  \bibfield  {author} {\bibinfo {author} {\bibfnamefont {P.~T.}\ \bibnamefont
  {Rakich}}, \bibinfo {author} {\bibfnamefont {C.}~\bibnamefont {Reinke}},
  \bibinfo {author} {\bibfnamefont {R.}~\bibnamefont {Camacho}}, \bibinfo
  {author} {\bibfnamefont {P.}~\bibnamefont {Davids}}, \ and\ \bibinfo {author}
  {\bibfnamefont {Z.}~\bibnamefont {Wang}},\ }\href {\doibase
  10.1103/PhysRevX.2.011008} {\bibfield  {journal} {\bibinfo  {journal} {Phys.
  Rev. X}\ }\textbf {\bibinfo {volume} {2}},\ \bibinfo {pages} {011008}
  (\bibinfo {year} {2012})}\BibitemShut {NoStop}%
\bibitem [{\citenamefont {Kang}\ \emph {et~al.}(2009)\citenamefont {Kang},
  \citenamefont {Nazarkin}, \citenamefont {Brenn},\ and\ \citenamefont
  {Russell}}]{kangNP2009}%
  \BibitemOpen
  \bibfield  {author} {\bibinfo {author} {\bibfnamefont {M.~S.}\ \bibnamefont
  {Kang}}, \bibinfo {author} {\bibfnamefont {A.}~\bibnamefont {Nazarkin}},
  \bibinfo {author} {\bibfnamefont {A.}~\bibnamefont {Brenn}}, \ and\ \bibinfo
  {author} {\bibfnamefont {P.~St.~J.}\ \bibnamefont {Russell}},\ }\href
  {\doibase 10.1038/nphys1217} {\bibfield  {journal} {\bibinfo  {journal}
  {Nature Physics}\ }\textbf {\bibinfo {volume} {5}},\ \bibinfo {pages} {276}
  (\bibinfo {year} {2009})}\BibitemShut {NoStop}%
\bibitem [{\citenamefont {Elser}\ \emph {et~al.}(2006)\citenamefont {Elser},
  \citenamefont {Andersen}, \citenamefont {Korn}, \citenamefont {Gl\"ockl},
  \citenamefont {Lorenz}, \citenamefont {Marquardt},\ and\ \citenamefont
  {Leuchs}}]{elserPRL2006}%
  \BibitemOpen
  \bibfield  {author} {\bibinfo {author} {\bibfnamefont {D.}~\bibnamefont
  {Elser}}, \bibinfo {author} {\bibfnamefont {U.~L.}\ \bibnamefont {Andersen}},
  \bibinfo {author} {\bibfnamefont {A.}~\bibnamefont {Korn}}, \bibinfo {author}
  {\bibfnamefont {O.}~\bibnamefont {Gl\"ockl}}, \bibinfo {author}
  {\bibfnamefont {S.}~\bibnamefont {Lorenz}}, \bibinfo {author} {\bibfnamefont
  {C.}~\bibnamefont {Marquardt}}, \ and\ \bibinfo {author} {\bibfnamefont
  {G.}~\bibnamefont {Leuchs}},\ }\href {\doibase 10.1103/PhysRevLett.97.133901}
  {\bibfield  {journal} {\bibinfo  {journal} {Phys. Rev. Lett.}\ }\textbf
  {\bibinfo {volume} {97}},\ \bibinfo {pages} {133901} (\bibinfo {year}
  {2006})}\BibitemShut {NoStop}%
\bibitem [{\citenamefont {Shelby}\ \emph {et~al.}(1985)\citenamefont {Shelby},
  \citenamefont {Levenson},\ and\ \citenamefont {Bayer}}]{shelbyPRB1985}%
  \BibitemOpen
  \bibfield  {author} {\bibinfo {author} {\bibfnamefont {R.~M.}\ \bibnamefont
  {Shelby}}, \bibinfo {author} {\bibfnamefont {M.~D.}\ \bibnamefont
  {Levenson}}, \ and\ \bibinfo {author} {\bibfnamefont {P.~W.}\ \bibnamefont
  {Bayer}},\ }\href {\doibase 10.1103/PhysRevB.31.5244} {\bibfield  {journal}
  {\bibinfo  {journal} {Phys. Rev. B}\ }\textbf {\bibinfo {volume} {31}},\
  \bibinfo {pages} {5244} (\bibinfo {year} {1985})}\BibitemShut {NoStop}%
\bibitem [{\citenamefont {Dainese}\ \emph
  {et~al.}(2006{\natexlab{a}})\citenamefont {Dainese}, \citenamefont {Russell},
  \citenamefont {Joly}, \citenamefont {Knight}, \citenamefont {Wiederhecker},
  \citenamefont {Fragnito}, \citenamefont {Laude},\ and\ \citenamefont
  {Khelif}}]{daineseNP2006}%
  \BibitemOpen
  \bibfield  {author} {\bibinfo {author} {\bibfnamefont {P.}~\bibnamefont
  {Dainese}}, \bibinfo {author} {\bibfnamefont {P.~St.~J.}\ \bibnamefont
  {Russell}}, \bibinfo {author} {\bibfnamefont {N.}~\bibnamefont {Joly}},
  \bibinfo {author} {\bibfnamefont {J.}~\bibnamefont {Knight}}, \bibinfo
  {author} {\bibfnamefont {G.}~\bibnamefont {Wiederhecker}}, \bibinfo {author}
  {\bibfnamefont {H.}~\bibnamefont {Fragnito}}, \bibinfo {author}
  {\bibfnamefont {V.}~\bibnamefont {Laude}}, \ and\ \bibinfo {author}
  {\bibfnamefont {A.}~\bibnamefont {Khelif}},\ }\href {\doibase
  10.1038/nphys315} {\bibfield  {journal} {\bibinfo  {journal} {Nat. Phys.}\
  }\textbf {\bibinfo {volume} {2}},\ \bibinfo {pages} {388} (\bibinfo {year}
  {2006}{\natexlab{a}})}\BibitemShut {NoStop}%
\bibitem [{\citenamefont {Beugnot}\ \emph
  {et~al.}(2007{\natexlab{a}})\citenamefont {Beugnot}, \citenamefont
  {Sylvestre}, \citenamefont {Maillotte}, \citenamefont {M\'{e}lin},\ and\
  \citenamefont {Laude}}]{beugnotOL2007}%
  \BibitemOpen
  \bibfield  {author} {\bibinfo {author} {\bibfnamefont {J.-C.}\ \bibnamefont
  {Beugnot}}, \bibinfo {author} {\bibfnamefont {T.}~\bibnamefont {Sylvestre}},
  \bibinfo {author} {\bibfnamefont {H.}~\bibnamefont {Maillotte}}, \bibinfo
  {author} {\bibfnamefont {G.}~\bibnamefont {M\'{e}lin}}, \ and\ \bibinfo
  {author} {\bibfnamefont {V.}~\bibnamefont {Laude}},\ }\href {\doibase
  10.1364/OL.32.000017} {\bibfield  {journal} {\bibinfo  {journal} {Opt.
  Lett.}\ }\textbf {\bibinfo {volume} {32}},\ \bibinfo {pages} {17} (\bibinfo
  {year} {2007}{\natexlab{a}})}\BibitemShut {NoStop}%
\bibitem [{\citenamefont {Shen}\ and\ \citenamefont
  {Bloembergen}(1965)}]{shenPR1965}%
  \BibitemOpen
  \bibfield  {author} {\bibinfo {author} {\bibfnamefont {Y.~R.}\ \bibnamefont
  {Shen}}\ and\ \bibinfo {author} {\bibfnamefont {N.}~\bibnamefont
  {Bloembergen}},\ }\href {\doibase 10.1103/PhysRev.137.A1787} {\bibfield
  {journal} {\bibinfo  {journal} {Phys. Rev.}\ }\textbf {\bibinfo {volume}
  {137}},\ \bibinfo {pages} {A 1787} (\bibinfo {year} {1965})}\BibitemShut
  {NoStop}%
\bibitem [{\citenamefont {Kroll}(1965)}]{krollJAP1965}%
  \BibitemOpen
  \bibfield  {author} {\bibinfo {author} {\bibfnamefont {N.~M.}\ \bibnamefont
  {Kroll}},\ }\href {\doibase 10.1063/1.1713918} {\bibfield  {journal}
  {\bibinfo  {journal} {J. Appl. Phys.}\ }\textbf {\bibinfo {volume} {36}},\
  \bibinfo {pages} {34} (\bibinfo {year} {1965})}\BibitemShut {NoStop}%
\bibitem [{\citenamefont {Laude}\ \emph {et~al.}(2005)\citenamefont {Laude},
  \citenamefont {Khelif}, \citenamefont {Benchabane}, \citenamefont {Wilm},
  \citenamefont {Sylvestre}, \citenamefont {Kibler}, \citenamefont {Mussot},
  \citenamefont {Dudley},\ and\ \citenamefont {Maillotte}}]{laudePRB2005}%
  \BibitemOpen
  \bibfield  {author} {\bibinfo {author} {\bibfnamefont {V.}~\bibnamefont
  {Laude}}, \bibinfo {author} {\bibfnamefont {A.}~\bibnamefont {Khelif}},
  \bibinfo {author} {\bibfnamefont {S.}~\bibnamefont {Benchabane}}, \bibinfo
  {author} {\bibfnamefont {M.}~\bibnamefont {Wilm}}, \bibinfo {author}
  {\bibfnamefont {T.}~\bibnamefont {Sylvestre}}, \bibinfo {author}
  {\bibfnamefont {B.}~\bibnamefont {Kibler}}, \bibinfo {author} {\bibfnamefont
  {A.}~\bibnamefont {Mussot}}, \bibinfo {author} {\bibfnamefont {J.~M.}\
  \bibnamefont {Dudley}}, \ and\ \bibinfo {author} {\bibfnamefont
  {H.}~\bibnamefont {Maillotte}},\ }\href {\doibase 10.1103/PhysRevB.71.045107}
  {\bibfield  {journal} {\bibinfo  {journal} {Phys. Rev. B}\ }\textbf {\bibinfo
  {volume} {71}},\ \bibinfo {pages} {045107} (\bibinfo {year}
  {2005})}\BibitemShut {NoStop}%
\bibitem [{\citenamefont {Beugnot}\ \emph
  {et~al.}(2007{\natexlab{b}})\citenamefont {Beugnot}, \citenamefont
  {Sylvestre}, \citenamefont {Alasia}, \citenamefont {Maillotte}, \citenamefont
  {Laude}, \citenamefont {Monteville}, \citenamefont {Provino}, \citenamefont
  {Traynor}, \citenamefont {Mafang},\ and\ \citenamefont
  {Th\'evenaz}}]{beugnotOE2007}%
  \BibitemOpen
  \bibfield  {author} {\bibinfo {author} {\bibfnamefont {J.~C.}\ \bibnamefont
  {Beugnot}}, \bibinfo {author} {\bibfnamefont {T.}~\bibnamefont {Sylvestre}},
  \bibinfo {author} {\bibfnamefont {D.}~\bibnamefont {Alasia}}, \bibinfo
  {author} {\bibfnamefont {H.}~\bibnamefont {Maillotte}}, \bibinfo {author}
  {\bibfnamefont {V.}~\bibnamefont {Laude}}, \bibinfo {author} {\bibfnamefont
  {A.}~\bibnamefont {Monteville}}, \bibinfo {author} {\bibfnamefont
  {L.}~\bibnamefont {Provino}}, \bibinfo {author} {\bibfnamefont
  {N.}~\bibnamefont {Traynor}}, \bibinfo {author} {\bibfnamefont {S.~F.}\
  \bibnamefont {Mafang}}, \ and\ \bibinfo {author} {\bibfnamefont
  {L.}~\bibnamefont {Th\'evenaz}},\ }\href {\doibase 10.1364/OE.15.015517}
  {\bibfield  {journal} {\bibinfo  {journal} {Opt. Express}\ }\textbf {\bibinfo
  {volume} {15}},\ \bibinfo {pages} {15517} (\bibinfo {year}
  {2007}{\natexlab{b}})}\BibitemShut {NoStop}%
\bibitem [{\citenamefont {Royer}\ and\ \citenamefont
  {Dieulesaint}(1999)}]{royerBOOK1999}%
  \BibitemOpen
  \bibfield  {author} {\bibinfo {author} {\bibfnamefont {D.}~\bibnamefont
  {Royer}}\ and\ \bibinfo {author} {\bibfnamefont {E.}~\bibnamefont
  {Dieulesaint}},\ }\href@noop {} {\emph {\bibinfo {title} {Elastic waves in
  solids}}}\ (\bibinfo  {publisher} {Wiley},\ \bibinfo {address} {New York},\
  \bibinfo {year} {1999})\BibitemShut {NoStop}%
\bibitem [{\citenamefont {Koyamada}\ \emph {et~al.}(2004)\citenamefont
  {Koyamada}, \citenamefont {Sato}, \citenamefont {Nakamura}, \citenamefont
  {Sotobayashi},\ and\ \citenamefont {Chujo}}]{koyamadaJOLT2004}%
  \BibitemOpen
  \bibfield  {author} {\bibinfo {author} {\bibfnamefont {Y.}~\bibnamefont
  {Koyamada}}, \bibinfo {author} {\bibfnamefont {S.}~\bibnamefont {Sato}},
  \bibinfo {author} {\bibfnamefont {S.}~\bibnamefont {Nakamura}}, \bibinfo
  {author} {\bibfnamefont {H.}~\bibnamefont {Sotobayashi}}, \ and\ \bibinfo
  {author} {\bibfnamefont {W.}~\bibnamefont {Chujo}},\ }\href
  {http://jlt.osa.org/abstract.cfm?URI=jlt-22-2-631} {\bibfield  {journal}
  {\bibinfo  {journal} {J. Lightwave Technol.}\ }\textbf {\bibinfo {volume}
  {22}},\ \bibinfo {pages} {631} (\bibinfo {year} {2004})}\BibitemShut
  {NoStop}%
\bibitem [{\citenamefont {Moiseyenko}\ and\ \citenamefont
  {Laude}(2011)}]{moiseyenkoPRB2011}%
  \BibitemOpen
  \bibfield  {author} {\bibinfo {author} {\bibfnamefont {R.~P.}\ \bibnamefont
  {Moiseyenko}}\ and\ \bibinfo {author} {\bibfnamefont {V.}~\bibnamefont
  {Laude}},\ }\href {\doibase 10.1103/PhysRevB.83.064301} {\bibfield  {journal}
  {\bibinfo  {journal} {Phys. Rev. B}\ }\textbf {\bibinfo {volume} {83}},\
  \bibinfo {pages} {064301} (\bibinfo {year} {2011})}\BibitemShut {NoStop}%
\bibitem [{\citenamefont {Thomas}\ \emph {et~al.}(1979)\citenamefont {Thomas},
  \citenamefont {Rowell}, \citenamefont {van Driel},\ and\ \citenamefont
  {Stegeman}}]{ThomasPRB1979}%
  \BibitemOpen
  \bibfield  {author} {\bibinfo {author} {\bibfnamefont {P.~J.}\ \bibnamefont
  {Thomas}}, \bibinfo {author} {\bibfnamefont {N.~L.}\ \bibnamefont {Rowell}},
  \bibinfo {author} {\bibfnamefont {H.~M.}\ \bibnamefont {van Driel}}, \ and\
  \bibinfo {author} {\bibfnamefont {G.~I.}\ \bibnamefont {Stegeman}},\ }\href
  {\doibase 10.1103/PhysRevB.19.4986} {\bibfield  {journal} {\bibinfo
  {journal} {Phys. Rev. B}\ }\textbf {\bibinfo {volume} {19}},\ \bibinfo
  {pages} {4986} (\bibinfo {year} {1979})}\BibitemShut {NoStop}%
\bibitem [{\citenamefont {Dainese}\ \emph
  {et~al.}(2006{\natexlab{b}})\citenamefont {Dainese}, \citenamefont {Russell},
  \citenamefont {Wiederhecker}, \citenamefont {Joly}, \citenamefont {Fragnito},
  \citenamefont {Laude},\ and\ \citenamefont {Khelif}}]{daineseOE2006}%
  \BibitemOpen
  \bibfield  {author} {\bibinfo {author} {\bibfnamefont {P.}~\bibnamefont
  {Dainese}}, \bibinfo {author} {\bibfnamefont {P.~St.~J.}\ \bibnamefont
  {Russell}}, \bibinfo {author} {\bibfnamefont {G.~S.}\ \bibnamefont
  {Wiederhecker}}, \bibinfo {author} {\bibfnamefont {N.}~\bibnamefont {Joly}},
  \bibinfo {author} {\bibfnamefont {H.~L.}\ \bibnamefont {Fragnito}}, \bibinfo
  {author} {\bibfnamefont {V.}~\bibnamefont {Laude}}, \ and\ \bibinfo {author}
  {\bibfnamefont {A.}~\bibnamefont {Khelif}},\ }\href {\doibase
  10.1364/OE.14.004141} {\bibfield  {journal} {\bibinfo  {journal} {Optics
  express}\ }\textbf {\bibinfo {volume} {14}},\ \bibinfo {pages} {4141}
  (\bibinfo {year} {2006}{\natexlab{b}})}\BibitemShut {NoStop}%
\bibitem [{\citenamefont {Wiederhecker}\ \emph {et~al.}(2008)\citenamefont
  {Wiederhecker}, \citenamefont {Brenn}, \citenamefont {Fragnito},\ and\
  \citenamefont {Russell}}]{wiederheckerPRL2008}%
  \BibitemOpen
  \bibfield  {author} {\bibinfo {author} {\bibfnamefont {G.~S.}\ \bibnamefont
  {Wiederhecker}}, \bibinfo {author} {\bibfnamefont {A.}~\bibnamefont {Brenn}},
  \bibinfo {author} {\bibfnamefont {H.~L.}\ \bibnamefont {Fragnito}}, \ and\
  \bibinfo {author} {\bibfnamefont {P.~St.~J.}\ \bibnamefont {Russell}},\ }\href
  {\doibase 10.1103/PhysRevLett.100.203903} {\bibfield  {journal} {\bibinfo
  {journal} {Physical review letters}\ }\textbf {\bibinfo {volume} {100}},\
  \bibinfo {pages} {203903} (\bibinfo {year} {2008})}\BibitemShut {NoStop}%
\bibitem [{Note1()}]{Note1}%
  \BibitemOpen
  \bibinfo {note} {The eigenvector is actually exactly a plane wave if the
  material constituting the waveguide is homogeneous, in which case Eq. (\ref
  {eq8}) is simply the Cristofell equation for elastic waves in
  solids}\BibitemShut {NoStop}%
\bibitem [{\citenamefont {Van~Thourhout}\ and\ \citenamefont
  {Roels}(2010)}]{vantourhoutNP2010}%
  \BibitemOpen
  \bibfield  {author} {\bibinfo {author} {\bibfnamefont {D.}~\bibnamefont
  {Van~Thourhout}}\ and\ \bibinfo {author} {\bibfnamefont {J.}~\bibnamefont
  {Roels}},\ }\href {\doibase 10.1038/nphoton.2010.72} {\bibfield  {journal}
  {\bibinfo  {journal} {Nature Photonics}\ }\textbf {\bibinfo {volume} {4}},\
  \bibinfo {pages} {211} (\bibinfo {year} {2010})}\BibitemShut {NoStop}%
\bibitem [{\citenamefont {Kippenberg}\ and\ \citenamefont
  {Vahala}(2008)}]{kippenbergS2008}%
  \BibitemOpen
  \bibfield  {author} {\bibinfo {author} {\bibfnamefont {T.}~\bibnamefont
  {Kippenberg}}\ and\ \bibinfo {author} {\bibfnamefont {K.}~\bibnamefont
  {Vahala}},\ }\href {\doibase 10.1126/science.1156032} {\bibfield  {journal}
  {\bibinfo  {journal} {Science}\ }\textbf {\bibinfo {volume} {321}},\ \bibinfo
  {pages} {1172} (\bibinfo {year} {2008})}\BibitemShut {NoStop}%
\end{thebibliography}%

\end{document}